\documentstyle[aps,prd]{revtex} 

\draft

\begin{document}
\twocolumn[\hsize\textwidth\columnwidth\hsize\csname
@twocolumnfalse\endcsname
\title{%
\hbox to\hsize{\normalsize\rm May 1998
\hfil Preprint MPI-PTh/98-40}
\vskip 36pt
Measurement Process In A Two-Barrier System
}
\author{L.~Stodolsky}
\address{Max-Planck-Institut f\"ur Physik 
(Werner-Heisenberg-Institut),
F\"ohringer Ring 6, 80805 M\"unchen, Germany}
\maketitle
\begin{abstract}
The description of a measuring process, such as that which occurs
when a quantum point contact
(QPC) detector is influenced by a nearby external electron which
can take up  two possible positions, provides a
interesting
application of the method of quantum
damping. We find a number of new effects,  due to the complete
treatment of phases afforded by the
formalism, although our results are generally similiar to those of
other treatments, particularly to those of Buks et al. 
 These are effects depending on the phase shift in the detector,
effects
 which depend on the
direction of the measuring current,  and 
 in addition to damping or dissipative effects, an
 energy shift of the measured system. In particular, the phase
shift effect leads to the conclusion  that
there can be effects of ``observation" even when the two barriers
in question pass the same current. 

  The nature of the  current through the barriers and its
statistics is discussed, giving a description of 
correlations in the current due to ``measurement" and the origin of
``telegraphic" signals.

\end{abstract}
\vskip2.0pc]


\section{Introduction}
 The measurement process using a  quantum point contact (QPC)
detector~\cite{buks}
can be described as the modification of a barrier whose
transmission
varies~\cite{field} according to whether an external electron is
nearby 
or further away.
When the external electron is close by there is a certain higher 
barrier,
and when it 
is further away, there is a reduced barrier. Given an incident 
or probing flux on the barrier, 
 the modification of the resulting current through the barrier ,
which then can be observed by
conventional means, thus ``measures"
where the external electron is located.

Experiments of this type give a fundamental insight into the nature
of measurement.
In an elegant  recent  experiment  Buks et al.~\cite{buks},-
stimulated by 
the work of Gurvitz~\cite{sug}- saw the expected loss of fringe
contrast
in an electron interference arrangement when  one of the paths in
the
interferometer was ``under observation" by a QPC ~\cite{al}.

  This effect is due to  the ``damping" or
``decoherence" arising from the creation of correlations between
the position of the electron in the interferometer and the
coordinates of the ``environment" or  ``observer "~\cite{us,ab}. A
number of
theoretical treatments of such measurement
processes have
been given in
 recent years. 
 Here we would
like to use the ``quantum damping" method of refs \cite{us,ab},
which
was devised to deal with precisely such questions and which
  gives a complete and transparent
treatment of the two-barrier  process. Furthermore it yields
 a number
of new results. These
 include the effects of 
 a phase which although it has no effect on the measuring current
nevertheless
 contributes to the  damping or decoherence, effects
 which depend on the direction of the detector current, and an
energy
shift of the 
observed system   induced  by the observing  process. 

 To elucidate the method we first consider a simple and well
understood
 situation, the two-state system. We shall  consider an
external electron which can be in one of two states and which at
the
same time  is ``observed" by a QPC or similar detector involving 
two barriers.
 A physical realization of the two-state system could be provided
 by two
 adjoining quantum dots. The  electron
can then be on either one of the  two dots, ``left'' or ``right''
 and  tunnel between them. 
( This tunneling for the external
electron
 should not be confused with that associated with
 the two
barriers of  the detector). The detector barriers will be higher or
lower
according
to whether the external electron is on the nearby or distant dot,
thus furnishing
a reading of the position of the external electron.

The focus of our attention is
 the $2x2$ density matrix $\rho$
for  the two-state system and in particular the effect the presence
of the detector has on the time
development of $\rho$. 
We should perhaps stress that, while the study of this
 density matrix leads to an understanding of the evolution  of the
state of the measured system,
(the electron
on the dots)
it does not of itself lead to an immediate
 understanding of the current in the detector
circuit. Thus an understanding of the nature of that
current must await
some further
steps (see below).

 The density matrix is characterized  by a ``polarization
vector'' $\bf P$, via $\rho={1\over 2} (I+\bf{P\cdot\sigma})$,
where
the  $\bf \sigma$ are 
the pauli matrices.
  $P_z$ gives the
probability for finding  the electron on the left or right dot via
$P_z=
Prob(L)-Prob(R)$.
while the other components of  $\bf P$  contain  information
characterizing 
the nature of the coherence.  More precisely
 $\vert {\bf P} \vert=1 $ means the system is in a pure state,
 while $\vert {\bf P} \vert=0 $
means that it is completely randomized or ``decohered".  
  $\bf P$ will both rotate in time
due to the real energies in the problem and shrink in length due to
the
damping or decoherence.
 The time development of $\bf P$  is
given by a ``Bloch-like" equation~\cite{us,ab}

\begin{equation}\label{pdot}
{\bf \dot P} = {\bf V x P} - D {\bf P}_{tr} 
\end{equation}

The three real energies $\bf V$ have the following significance in
the present
problem, where the two-dot system may be thought of as a double
potential
well for the external electron.  
  $V_z$ gives a possible energy difference for
the two quasi-stationary states on each dot; $V_z\neq 0$ means the
the double well is asymmetric. $V_x $ and $V_y$ are 
tunneling energies; $V_x$ conserves the parity of the electron
wavefunction
on the
two dots and $V_y$ flips it~\cite{cina}.

The second term of Eq.~(\ref{pdot}) gives the damping or
decoherence. $D$
 gives the rate  at which
correlations 
are being created between the ``system " (the external electron on
the dots)
and the ``environment" (the detector).
 
The label ``tr" on ${\bf P}_{tr}$  means ``transverse" to the $z$
axis.
 The ``z" direction is selected  by the fact
 that
the measuring process does not cause the  electron to jump from one
dot
to
another; the measuring process conserves $P_z=<\sigma_z>$, and so
only
damps the transverse components of
$\bf P$. (It should be recalled~\cite{ab} that a measurement must
choose
 some axis or  direction in the hilbert space, otherwise there is
no measurement 
at all.)

\section{\bf Influence of the Detector }

 The effect
of
 repeated probings by an ``environment" or ``measuring device" 
is described by a quantity $\Lambda$ whose imaginary part gives the
damping and whose real part gives an  energy shift to the 
system being measured. Thus we expect that the observation
process here
will cause  not only a damping or  dissipation but also an 
energy shift for  the external electron
 on  the two dots. 

 $D$ is thus the imaginary part of a quantity $\Lambda$, while the
real part of $\Lambda$ leads to a observation-induced contribution
to $\bf V$. $\Lambda$ itself is given by

\begin{equation}\label{lam}
\Lambda= i (flux)<i\vert{ 1-S_LS_R^\dagger  }\vert i>
\end{equation}

The factor $flux$ is the flux or probing rate of the detector
electrons,
where in the QPC application one
can use the  Landauer
formula $flux = eV_d/\pi\hbar$,
with $V_d$ the voltage in the detector circuit~\cite{thou}). The
label $i$
refers to
the 
the initial or incoming state of the electrons in the detector
and the $S$'s are the S matrices for the two barriers 
created by the  
 two locations of the external electron.  

{\it S Matrix-}\quad In order to apply Eq.~(\ref{lam}) , we review
shortly
how to express the barrier penetration  problem in S-matrix
language.
 At a given energy there are just
two S-matrix
elements for the incoming flux with wave vector $k$, namely
$S_{kk}$ and $S_{-kk}$, representing transmission and reflection
respectively.
 These are the coefficients appearing in the wavefunction of unit
incoming amplitude
 which  far away from the barrier on the incoming side
is
\begin{equation}\label{wav}
 e^{ikz}+S_{-kk}e^{-ikz}
\end{equation}
while on the other, outgoing, side of the barrier at large
distances we have
$ S_{kk}e^{ikz}.$

 We may also have the incoming detector current  from the other 
direction
 with wave vector $-k$ so
 we have $S_{-k-k}$ and $S_{k-k}$ for transmission and
reflection.
 Thus our S-matrix is
$S=\pmatrix{S_{kk}&S_{k-k}\cr S_{-kk}&S_{-k-k}}$
where the two columns correspond to the two possible directions
of the detector current.
 This matrix is unitary, $SS^\dagger=1$, as may be found from
explicit constructions.

{\it Time Reversal-}\quad  An additional constraint arises 
from time reversal invariance. This
 states  that
$ S_{j,i}=S_{i_T,j_T}, $ where
the subscript $T$ means the time-reversed state.  Here we have
simply
 $k_T=-k$ and vice-versa.
For the diagonal elements this gives 
$ S_{kk}=S_{-k-k}$
while for the off-diagonal elements their is no further constraint
beyond that  already given by unitarity.
Note that we neglect  electrons  spin and there is no
magnetic 
field; these further complications
might be interesting in some applications.

We can now parameterize the S-matrix with 
the time reversal constraint  in terms of three
angles as
\begin{equation}\label{sma}
 S=e^{i\phi}\pmatrix{ cos\theta &ie^{-i\eta}sin\theta\cr
 ie^{i\eta}sin\theta & cos\theta} 
\end{equation}
which it will be seen fulfills all the conditions just discussed.
The first column contains the reflection and transmission
coefficients for incoming waves $k$ and the second column those for
incoming waves with $-k$.
The angle $\theta$, which gives the magnitude of transmission
and reflection, is the same as that used by Buks et al, while the
other two parameters are phases.

  The phase angle $\eta$ creates
a difference between incident waves with $k$ and $-k$, that is for
different 
directions of the detector current. This reflects
a possible asymmetry in the shape of the barrier (see below) and
leads, in the case of non-zero $\eta$,
 to the interesting possibility of effects which depend
 on the  direction of the measuring current. 

 We note that although $\phi$ appears as an overall phase it is 
physically relevant. The phases are fixed by
 the $``1"$ in Eq.~(\ref{lam}), or correspondingly by the fact that
we
have
 1 as the coefficient of the incoming wave in Eq.~(\ref{wav}).
 (Recall that in three-dimensional partial wave scattering theory
the
entire scattering
is given by   just a phase, $S_l=e^{2i\delta_l}$).

{\it Parity-}\quad If the barriers in question are even in shape,
 another constraint arises due to parity symmetry. When this
operation, namely
  $z\rightarrow -z$, is applied to the wavefunction Eq~(\ref{wav}),
we
  get a solution corresponding to a wave coming in from the other
direction.
Comparing coefficients we conclude
that $ S_{-k-k}=S_{kk}$ and $ S_{k-k}=S_{-kk}$. The first
condition was
 already obtained from time reversal, the second, however, says
that for
  symmetric barriers  we should set $\eta=0$ in Eq~(\ref{sma})
giving that the $S$ elements are the same for both directions
of the detector current. Note that a non-zero $\eta$ only affects
 the reflection coefficients;  even with  non-symmetric
barriers T ensures that the transmission coefficients have the same
phase.

\section{\bf Damping}

  $D$ is the parameter giving  the damping or loss of coherence
of the system, and
is given  by $D= Im \Lambda $. We label  
the parameters of the $S$ matrices  with an L or R for the
 two  barriers created when the electron is on the right dot or the
left
dot and call $\phi_L-\phi_R =\Delta \phi$,
 $\eta_L-\eta_R =\Delta \eta$ and $\theta_L-\theta_R =\Delta
\theta$.
We then have from the matrix ~(\ref{sma}),

\begin{equation}\label{dam}
 D= (flux)Re\big\{1-e^{i\Delta \phi}[cos \Delta \theta+
sin\theta_L sin\theta_R(e^{i\Delta \eta}-1)]\big\}
 \end{equation}

This applies for a given direction of the detector current, for the
other direction,
we reverse the sign of $\Delta \eta$.
For a symmetric barrier, this simplifies to
\begin{equation}\label{dama}
 D= (flux)\big\{1-cos\Delta \phi~ cos \Delta \theta
 \big\}
\end{equation}
 Except for the  $cos\Delta \phi$ factor this is essentially the
same damping effect as  in  ref~\cite{buks}.
This factor is interesting, however, in that
 even  for small angles,
\begin{equation}\label{damb}
D\approx 1/2 (flux)\big\{(\Delta \phi)^2+  (\Delta \theta)^2
 \big\}
  \end{equation}
the phase $\phi$ is present and enters on an equal footing with 
$\theta$.
Since according to the present method all phases should be kept,
$\phi$
is meaningful and contributes to the damping or
decoherence~\cite{pha}.

 This says
that two detector  barriers which have the same transmissibility,
that is the same $\theta$,
but different phase shifts can nevertheless induce damping. 
Although they
apparently
give the same detector current they nevertheless establish a
distinction
or ``make
a measurement". This may seem less mysterious when we recall that
even pure 
phase shifts correspond to physical effects like the delay or
change in shape
of wave packets. Although a change in current is the most obvious
``measurement",
it is not necessarily the only one. Thus we
apparently differ with  other
treatments ~\cite{buks},~\cite{al},\cite{gr2},\cite{lev}, where
the damping or decoherence is related to the detector
current only.
Our results
are of course proportional to the flux or probing rate, but not
necessarily to the
transmitted  current alone. In effect, the situation concerning
$\phi$
may be
viewed as
 case of the well-known tree falling in the forest with nobody
there to hear it.
\section{\bf Energy Shift}

 While the damping is given by the imaginary part of
Eq~(\ref{lam}), there is
also a 
significance to the real part. In the description of 
the 
propagation of a particle in a medium it gives the index of
refraction
for the states
of the particle in the medium~\cite{ab}. In the present problem
this state-dependent energy shift
 will give a measurement-induced contribution to ${\bf
V}_z$
 in Eq~(\ref{pdot}) governing the internal evolution of the
measured
system~\cite{ab},  ${\bf V}_z^{ind}= Re \Lambda$ so that

\begin{equation}\label{rea}
 V_z^{ind} = (flux)Im~ e^{i\Delta \phi}[cos\Delta
\theta+
sin\theta_L sin\theta_R(e^{i\Delta \eta}-1)]    
\end{equation}
 which for an even barrier with $\Delta \eta=0$ simplifies to
\begin{equation}\label{reb}
V_z^{ind}=  (flux) sin\Delta\phi~ cos \Delta\theta
\end{equation}

This effect only contributes to the ``z direction" of $\bf V$ for
the same
reason $D$ only effects ${\bf P}_{tr}$, the
observation process
is presumed not to cause any jumps from one dot to the other.
 We note that the effect persists even if $\Delta
\theta=0$, i.e. equal transmissibilities,
as long as $\Delta \phi \neq 0$; and for small phases that it is
linear in
$\Delta \phi$ while the damping Eq~ (\ref{damb}) is quadratic. 

This induced energy offers the intriguing opportunity of ``tuning"
the
properties of the two-state system. Since the tunneling behavior of
the external electron between the two dots depends strongly on how
exact
the degeneracy of the two wells is, the ability to adjust it 
via the induced $V_z$, which note is proportional to $V_d$ and so
easily adjustable, is quite interesting. For example, if it is
difficult to fabricate identical dots, the induced $V_z$ could be
used to nevertheless make the two dots degenerate in energy. 
 On the other hand a large $V_z $, by lifting an
initial near-degeneracy  tends to
suppress  transitions, offering another qualitative test of the
theory.

\section{\bf Dependence on Current Direction }   

For barriers that are not symmetric in shape, so
$\eta$ can be non-zero, the damping and the energy shift will in
general
depend on the direction of the detector current. To find the
difference
we exchange $k$ for $-k$ in the initial state. For the difference
in
$D$ for the
two directions we have

\begin{equation}\label{dif}
\Delta D =2(flux) sin\Delta\phi sin\theta_L sin\theta_R
 sin\Delta\eta
\end{equation}
and for the induced energy shift the difference for the two
directions

\begin{equation}\label{difa}
\Delta V_z =2(flux) cos\Delta\phi sin\theta_L sin\theta_R
 sin\Delta\eta
\end{equation}

Both these effects are linear for small $\Delta \eta$.

\section{\bf  Nature of the  Current }

 The question of the current in the detector circuit poses some
intriguing questions
concerning the nature of ``measurement".   Should we expect a
smooth current of  some kind, reflecting some  average transmission
probablity given by $\rho$? Or should we view each transmission as
a measurement which 
``collapses  the wavefunction''  to one dot or the other
leading to
a series of
``telegraphic" signals with different currents  corresponding to
one  barrier  or  the other? If so, what determines the duration of
these signals?
   
The above considerations for the determination of $\rho$  cannot
alone answer such questions.  From
$\rho$ we can only find the probability of a single
transmission at a given time,  (the unconditional
probability)  but not if this event was say part
of a long series of transmissions(conditional probability).
These questions arise because
successive probings are in general not statistically independent;
one transmission may imply an increased probability for the next
one. A helpful analogy here might be  successive
Stern-Gerlach procedures: if a spin was ``up" passing through a
first
magnet, it will also be ``up" passing through a second magnet.

This problem of  finding  the probabilities for various sequences
of transmissions and reflections is perhaps most easily
addressed in an amplitude formulation~\cite{boris}. Let  1
represent  a transmission and  0
a reflection. We write the probability
amplitude
 that after N probings the external electron is on dot L, and that
  the first probing electron was transmitted, the second
reflected, ...
 the $N-1$th transmitted and 
the Nth transmitted,  as  $A(L,[11...01])$. Similarly there are
amplitudes
 $A(R,[.....])$ for the external electron to be on dot R after  a
given sequence of transmissions
 and reflections [.....].

 Let us now make the assumption that many probing electrons are
incident during the time $1/V_{tr}$ it takes the external  electron
to tunnel to the other dot:  
\begin{equation}\label{dl}
 flux / V_{tr} >> 1 
\end{equation}

This assumption allows us, in
discussing the current, 
to neglect ``dot jumps"  where
an amplitude $A(L,[...])$ receives a contribution from an
amplitude $A(R,[...])$. Then for time periods $\delta t$ such that
$V_{tr} \delta t<<1$ we can simply write

\begin{equation}\label{amp}
 A(L,[11...01])\sim S^L_{kk}S^L_{kk}...S^L_{-kk}S^L_{kk}
\end{equation}
  where we have suppressed phase factors associated
with the time,
and also the probability of the starting configuration. If we
restore
the latter, in the form of $\rho_{LL}$ for the probability of
starting with L, and $\rho_{RR}$ for the probability of starting
with R, we find by squaring the amplitude the probability for a
given sequence of transmissions and reflections:

\begin{equation}\label{pro}
Prob[11...01]=\rho_{LL}(p_Lp_L...q_Lp_L)+\rho_{RR}(p_Rp_R...q_Rp_R)
\end{equation}
 where $p=\vert S_{kk}\vert ^2 $ and $q$ is the probability of no
transmission, $q=1-p=\vert S_{-kk}\vert ^2$. (Note these quantities
are independent of
the current direction $k$, due to T invariance).

  Eq~(\ref{pro}) shows how a tendency to  ``telegraphic"
behavior with sequences of reflections or sequences of
transmissions can arise. For example, let $p_L$ and $p_R$ be very
different, say
close to zero and  close to one respectively. Then neither the
first term nor the second term can be big for mixed sequences like
[010...01]; but for ``telegraphic" sequences [111...11] or
[000...00], 
 one or the other term can be big. On the other hand if $p_L$ and
$p_R$ are very similar, this tendency favoring repeated signals
will be weak; little correlation between the ``measured" and the
``measurer" is introduced. 

 The two terms of
Eq~(\ref{pro})  are essentially those leading to 
the binomial distribution in statistics. Hence if we now ask for
the
probability $Prob(Q,N)$ for Q transmissions in N probings, the
combinatorics are that
of the binomial distribution, and we obtain
\begin{equation}\label{proba}
Prob(Q,N)=\rho_{LL}{\cal P}_L(Q)+\rho_{RR}{\cal P}_R(Q)
\end{equation}
 where  ${\cal P}_L(Q)$ is the binomial
expression for the probability of Q transmissions in N trials given
the single trial
probability
$p_L$. (For N large and p small this is approximated by the poisson
distribution ${\cal P}_L(Q)\approx {(\bar n_L)^Q \over Q!} e^{-\bar
n_L}$, with $\bar n_L=p_LN$).

 Eq (\ref {proba}), for distinct $p_L$ and $p_R$, leads to a two-
peaked distribution, and generally describes the statistics of the
current for short times. For example
the quantity $Prob(Q_1,Q_2)$ for $Q_1$ transmissions in $N_1$
probings, followed by $Q_2$ transmissions in $N_2$ probings gives
a measure of the
correlations induced by the ``measurement" process. In particular
$Prob(Q_1,Q_2)-Prob(Q_1)Prob(Q_2)$  is
zero if
the  currents in the two intervals are independent. 
Prob$(Q_1,Q_2)$ may be calculated as simply the
weighted average of two parallel processes, each one calculated as
a statistically independent sequence: Prob$(Q_1,Q_2)=\rho_{LL}
{\cal P}_L(Q_1){\cal P}_L(Q_2)+\rho_{RR}
{\cal P}_R(Q_1){\cal P}_R(Q_2)$, while Prob(Q) is given by ~Eq
(\ref{proba}). Thus 
\begin{eqnarray}\label{probb}
&&Prob(Q_1,Q_2)-Prob(Q_1)Prob(Q_2)= \\ \nonumber
&&\rho_{LL}(1-\rho_{LL})({\cal
P}_L(Q_1)-{\cal P}_R(Q_1))({\cal P}_L(Q_2)-{\cal P}_R(Q_2))
\end{eqnarray}
We expect  this  correlation to  vanish when only one dot is
occupied, for then successive probings are independent; it is in
fact zero for $\rho_{LL}$ 
one or zero. Similarly it vanishes   when the current provides no
information on the state,   
 when $p_L$ and $p_R$ are equal or  ${\cal P}_L={\cal P}_R$. 
Eq (\ref{probb}) thus gives a characterization of the strength of
``collapse" or ``telegraphic" effects.

 These considerations hold, as said, for times short relative to
the tunneling time $1/V_{tr}$. For longer times an amplitude
$A(L,[...])$ receives  contributions from an
amplitude $A(R,[...])$, thus we anticipate a time scale for the
correlations or a duration of the ``telegraphic" signals on the
order of $1/V_{tr}$.

\section{Strong Damping}
  In the previous section the time scale was the tunneling time
$1/V_{tr}$. However, there is a another regime ~\cite{us} of
behavior,
although it may not be relevant in problems where the damping  is
relatively weak, as  when the two barriers differ little, but which
is
of interest in itself. This is the case of strong damping, when
$D/V_{tr}>>1$. In the limiting case of strong damping the 
amplitude for any configuration 
$A(L,[....])$ is reached by only one ``path" $[....]$ and the
situation
resembles a classical diffusion problem. The solutions of
~Eq(\ref{pdot}) show the ``Turing-Watched-Pot-Zeno"
behavior~\cite{us} where ``measurement" inhibits the time evolution
and the characteristic time scale becomes the much longer
$D/V_{tr}^2$. In this case the condition~Eq (\ref{dl}) can be
weakened to $flux/(D/V_{tr}^2)>>1$, or since $D=flux$ in the limit,
 to $(flux/V_{tr})^2>>1$. Similarly the time scale for
the relaxation of correlations becomes the longer time
$D/V_{tr}^2$. 

\section{\bf Conclusions and Applications}

  If it were possible to fabricate the two dot system or its
analogs and
to carry out experiments with them, the various effects 
could be tested through their effect on  the diagonal
elements of $\rho$, or $P_z$. $P_z(t)$ gives the probability that
a dot is occupied, and through ~Eq(\ref{pdot}) it is influenced by
the real energies and $D$. One way of determining it experimentally
would be from the current described by ~Eq(\ref{proba}), that is
from the relative strengths of the two peaks of the $Q$
distribution.
It should
be stressed  that ~Eq(\ref{proba}) represents an average over many
repeated runs  from the same initial condition, say the injection
of the external electron onto dot L, and not an average over
time  in one run.

It should also be noted that even if the evolution of $\rho$ cannot
be followed in detail, the time scale of the relaxation of
correlations yields information on $V_{tr}$ and $D$.

 For experiments of the type of ref~\cite{buks}, 
the
real energy Eq~(\ref{rea}) induced by the measurement  will show
up as a shift in the interference fringes. In Eq~(\ref{rea}) the
effect
is expressed as
 an energy, that is as a phase per unit time, which in an
experiment
such as ref~\cite{buks} corresponds to a phase shift given by the
dwell time of the measured electron times the energy
Eq~(\ref{rea}).  
Hence Eq~(\ref{rea}) 
predicts, for non-zero $\phi$, a $V_d$-dependent fringe shift. For
the asymmetric barrier, Eq~(\ref{difa}) predicts  a
component to this fringe shift which reverses with $V_d$.
Similarly, for the asymmetric barrier, Eq~(\ref{dif}) predicts that
the loss of fringe contrast or visibility of ref~\cite{buks} can  
depend on the current direction. Most interesting 
 is the role of $\phi$. According to Eq~(\ref{damb}), if we
can arrange for the  two QPC barriers to have the same transmission
but different $\phi$
there still should be a loss of fringe contrast or damping.

  
I am grateful to the Weizmann quantum dot group for introducing me
to this subject and  especially to E. Buks, S. Gurvitz and M.
Heiblum for
several discussions, and to R. A. Harris for
 help in clarifying the ideas. Finally, I am grateful to  a
number of these colleagues, and in particular A. J. Leggett, for
criticisms concerning  the subject of section VI, finally leading
to the viewpoint developed there.  I would also like to acknowledge
the hospitality of the Institute of Advanced Studies at the Hebrew
University of Jerusalem, where this work was begun.



\end{document}